\documentclass[10pt,journal,compsoc]{IEEEtran}
\ifCLASSOPTIONcompsoc
  \usepackage[nocompress]{cite}
\else
  \usepackage{cite}
\fi
\ifCLASSINFOpdf
   \usepackage[pdftex]{graphicx}
\else
\fi
\usepackage{amsmath}
\usepackage{array}
\begin{document}
%
\title{vue4logs - Automatic Structuring of \\ Heterogeneous Computer System Logs}
%
%
%
%

\author{Isuru~Boyagane, Dept. of Computer Science and Engineering,              University of Moratuwa, Sri Lanka. \\
        Oshadha~Katulanda, Dept. of Computer Science and Engineering, University of Moratuwa, Sri Lanka. \\
        Surangika~Ranathunga, Senior Lecturer, Dept. of Computer Science and Engineering, Faculty of Engineering, University of Moratuwa, Sri Lanka. \\
        Srinath~Perera, WSO2 Inc.

}
%
%

\markboth{}%
{Shell \MakeLowercase{\textit{et al.}}: Bare Demo of IEEEtran.cls for Computer Society Journals}
%



\IEEEtitleabstractindextext{%
\begin{abstract}
Computer system log data is commonly used in system monitoring, performance characteristic investigation, workflow modeling and anomaly detection. Log data is inherently unstructured or semi-structured, which makes it harder to understand the event flow or other important information of a system by reading raw logs. The process of structuring log files first identifies the log message groups based on the system events that triggered them, and extracts an event template to represent the log messages of each event. This paper introduces a novel method to extract event templates from raw system log files, by using the vector space model commonly used in the field of Information Retrieval  to vectorize log data and group log messages into  event templates based on their vector similarity. Template extraction process is further enhanced with the use of character and length based filters. When evaluated on publicly available real-world log data benchmarks, this proposed method outperforms all the available state-of-the-art systems in terms of accuracy and robustness. NOTE: This work has been submitted to the IEEE for possible publication. Copyright may be transferred without notice, after which this version may no longer be accessible
\end{abstract}

}

\maketitle

\IEEEdisplaynontitleabstractindextext

%
\IEEEpeerreviewmaketitle

\IEEEraisesectionheading{\section{Introduction}\label{sec:introduction}}

%
%
%
%
\IEEEPARstart{C}{omputer} systems generate system log files that consist of run-time information of the system. A log file may contain a series of log lines that describes different system events occurred in the computer system. Each of these events in the log file originates from a respective code segment in the source code, which ensures that each longline originated from same code shares a underline pattern. Multiple occurrences of the same system event may generate multiple log lines of the event. The log lines that belong to a specific system event only differ from each other based on the dynamic states of the system event. 

\begin{figure}[!t]
\centering
\includegraphics[width=3.5in]{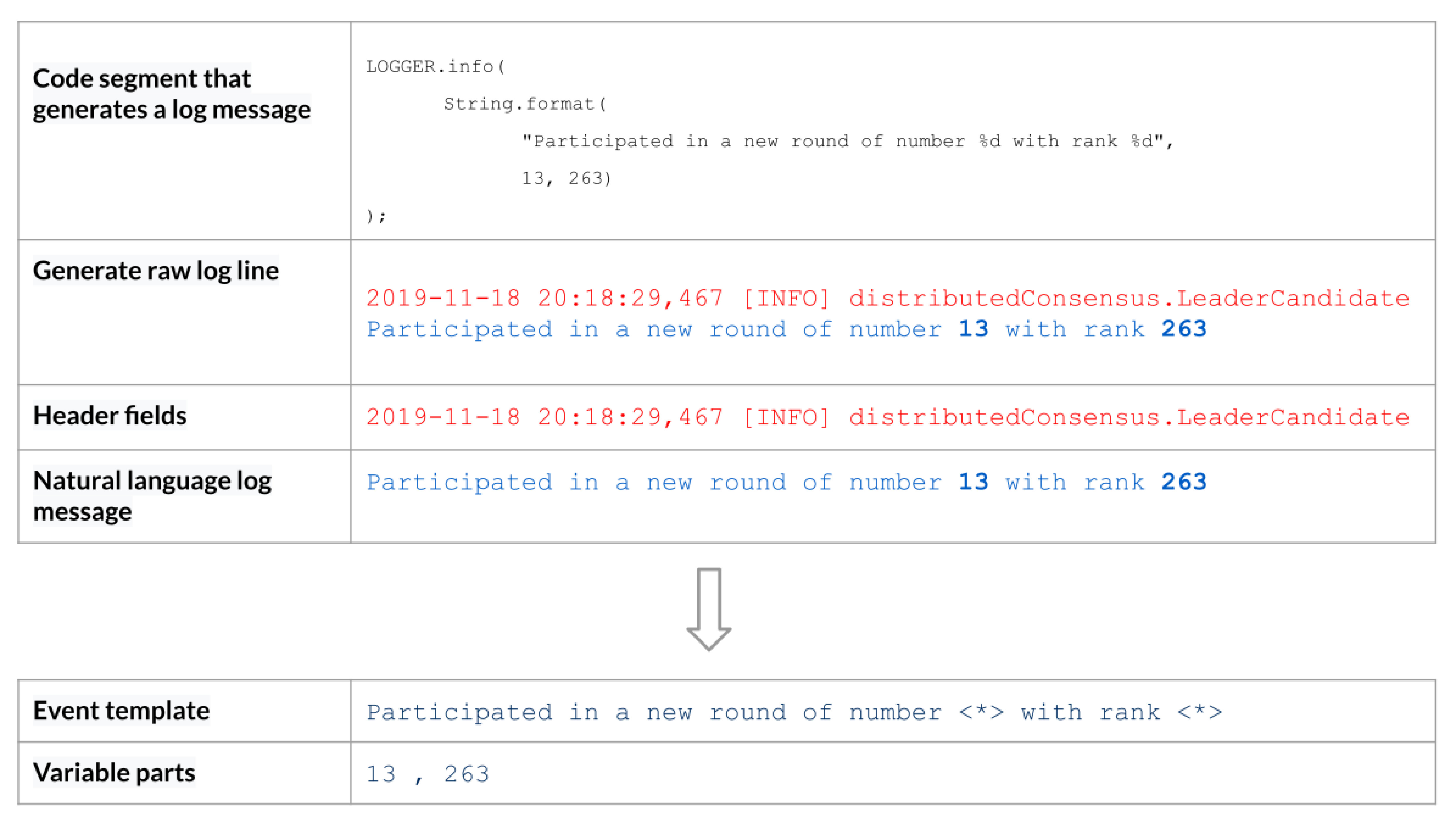}
\caption{System log message and the corresponding template.
}
\label{fig_sim}
\end{figure}

System generated log lines usually contain different header fields and a log message  (Fig. \ref{fig_sim}). The log message is written in natural language and carries detailed information about the system event that occurred.  The header fields are structured, and describe common information such as timestamp, verbosity level and component (Fig. \ref{fig_sim}). Thus, in the most general case, system logs can be considered semi-structured. 

The words present in a log message can be categorized into two basic types, namely \texttt{constant parts} and \texttt{variable parts} (Fig. \ref{fig_sim}). The \texttt{constant part} represents the event template or the event type of a log message and is invariant across the log messages that are generated by the same system event. In contrast, \texttt{variable parts} of a log message consist of dynamic information related to the particular event and can vary from one occurrence of the event to another. For instance, \texttt{263} and \texttt{13} in the log message shown in Fig. \ref{fig_sim} can be any number based on the dynamic state of the system at the logging time. In contrast, the \texttt{Participated in a new round of number ... with rank ...} part remains unchanged for any log message generated by the same event.

Common log analysis tasks include usage analysis, anomaly detection, performance modeling and failure diagnosis \cite{8804456}. However, directly using raw system log files as input to these analysis tasks is less effective due to the semi-structured nature of system logs. Thus, it is a common practice to structure these log files prior to using them in log analysis tasks, which requires tedious manual handling~\cite{8804456}. 

The process of automatically structuring the log messages is commonly identified as log parsing \cite{liu2020fastlogsim},\cite{8029742},\cite{7579781}, log signature extraction \cite{Thaler2017} or log template generation \cite{10.1145/2619287.2619290}. The term “log parsing” is used in this paper. In the process of log parsing, each of the log messages is assigned to its event group, and a textual template is extracted to represent all of the log messages of a particular event group.

Technically, the key objective of log parsing is to correctly identify the \texttt{constant part} and \texttt{variable parts} within the log messages \cite{8804456, 8029742, 8489912} and group log messages that have the same \texttt{constant part} together by assigning a representative event template for each event group. When extracting a representative template for the log messages generated by the same event, the positions of \texttt{variable parts} in the event template are replaced by a wildcard. Related literature has reported  “*”   \cite{7579781, 1251233} and "$<*>$" \cite{8804456, 9110435} as the most used wildcards. In this paper, "$<*>$" is used to represent the wildcard. Once extracted, the event template of a particular system event can be used as the structure of all of the log messages generated by a particular system event  (Fig. \ref{fig_sim}).

Most of the existing log parsing methods group log messages into events by observing the most general characteristics of log data \cite{8804456}. These characteristics are used in different stages of log structuring methods as heuristics. These methods have the potential to generate event templates with a very high accuracy for the data sets that comply with characteristics observed when designing the log parsing method. When new data violates the assumptions in the heuristic, the log parsing method tends to fail on such data. Applying general text clustering methods on log data is also used in this problem domain \cite{10.1145/2063576.2063690, 10.1145/2983323.2983358}. According to the Zhu et. al. \cite{8804456}, these text clustering methods do not perform well compared to the heuristic-based methods. Converting log messages into vector form and performing vector clustering techniques \cite{shima2016length} has performed reasonably \cite{8804456}.

Huang et. al. \cite{9110435} used Information Retrieval techniques for log parsing. Using publicly available benchmarks, they experimentally showed that it is the  best solution for log parsing with respect to the parsing accuracy. However, this method uses a Jaccard similarity-based technique to match the log messages against the identified event templates. This method is sub-optimal because Jaccard similarity focuses only on the text overlap between the target documents (log message and the template in our scenario). Jaccard similarity treats all words as equal and does not consider the importance of words to the documents. Thus it has less potential to capture the difference between the \texttt{constant parts} and \texttt{variable parts} in the log data. Furthermore, we can observed that some words present in the log data consist of slight character-level variations, which may lead to highly similar log messages to be considered as completely different when a word-level Jaccard similarity is calculated. 

To overcome these shortcomings, this paper presents a new technique named vue4logs. Instead of Jaccard Coefficient, vue4logs uses cosine similarity to measure the similarity of log messages and templates by means of vector similarity. Documents are converted to a vector representation using Term  Frequency - Document Frequency (TF-IDF), which is a numerical statistic that reflects how important a word is to a document in a collection \cite{Ramos2003UsingTT}.
Furthermore, vue4logs uses character-level text processing techniques (as discussed in section \ref{ext-mat-temp}) before creating the vocabulary to calculate TF-IDF, to overcome the challenges faced due to the presence of slight character-level variations inside the potential \texttt{constant} words. This technique, when used  along with the TF-IDF representation and cosine similarity, significantly improves the correctness of log structuring.

We evaluated vue4logs on sixteen publicly available datasets  \cite{8804456} that are commonly used as the benchmark data sets in related literature. The results show that it has the highest average parsing accuracy when compared to the state-of-the-art log parsing methods. Experiments show that this solution has the lowest variance of parsing accuracy across the evaluated sixteen log data sets, which is a sign of higher robustness. Furthermore, vue4logs and the same baseline systems were evaluated in a source-independent parameter tuning setting, which is more likely to be the case in a production-level system. The tuning configurations identified in this source-independent parameter tuning are considered as the data-independent tuning parameters of that particular parser and considered to be suitable to use on any custom data. In this evaluation setting also, vue4logs has shown the highest accuracy. We also evaluated vue4logs and the selected log parsing methods on a custom data set to investigate the usability of the methods on custom data. The accuracy results show that vue4logs performs better than the state-of-the-art methods on custom data sets as well. Thus vue4logs establishes a new baseline in log parsing. Our code is publicly released\footnote{https://github.com/IsuruBoyagane15/vue4logs-parser}. We also release a new annotated dataset that can be used to evaluate log parsing systems\footnote{https://github.com/IsuruBoyagane15/paddy-recreation/blob/main/ground\_truth/Forensic\_2k.log\_structured.csv}.

The rest of the paper is organized as follows. Section 2 discusses related work and Section 3 presents the vector space model for document similarity measurement. Section 4 presents the vue4logs system and Section 5 presents results. Finally, Section 6 concludes the paper.

\section{RELATED WORK}

Using rule-based techniques is the basic solution that is initially used to address the log parsing problem. Later, data-driven methods were introduced to make the log analysis process more efficient and less error-prone. This includes frequent pattern mining, text or vector based clustering techniques, heuristics-based techniques, deep learning techniques as well as Information Retrieval techniques. Other than these categorized techniques, more specific techniques such as the Longest Common Subsequence (LCS) are also used.


\subsection{Rule-based Techniques}

One of the basic approaches for log parsing is using handcrafted regular expressions in the form of grok patterns \cite{ABeginne19:online}. In this method, raw log files are manually analysed and regular expressions are formed to be used as the event templates. Once the regular expressions are formed, they can be used to extract the structure from priory unseen data. Although this method is straightforward and has been employed in production level systems as well \cite{Logstash57:online}, manually writing ad-hoc rules to structure large log files is time-consuming and error-prone.  Code bases in modern software systems get updated frequently, leading to an inevitable cost of regularly revising the set of regular expressions. 

\subsection{Frequent Pattern Mining}

Frequent pattern mining methods operate based on the assumption that the more frequent words of a log data are more likely to be \texttt{constant parts}. These methods basically traverse over the log data several times while identifying frequent item-sets (in most of the cases tokens or token position pairs), then cluster the log message based on the identified frequent items. Finally, an event template for each cluster is extracted \cite{8804456}. SLCT \cite{1251233} is the first method to use this technique on log parsing. LogCluster \cite{7367331} and LFA \cite{5463281} are the improved versions of SLCT \cite{8804456}. But, in a situations where the number of \texttt{constant parts} in infrequent event templates is less than that of \texttt{variable parts} in frequent event templates, these methods would consider the former as a \texttt{variable part} and fail to extract the correct event template. Therefore, log parsing techniques such as SLCT cannot precisely identify templates that are rare in the log dataset \cite{8804456, Thaler2017}. 

\subsection{Clustering}

Log parsing can be modelled as a clustering problem to identify the group of log messages that are generated by the same system event. LogSig \cite{10.1145/2063576.2063690} clustered log messages into a  predefined number of clusters \cite{8804456}. However, determining the number of clusters beforehand is difficult. LogMine \cite{10.1145/2983323.2983358} uses a bottom-to-top hierarchical clustering method to extract event templates \cite{8804456}. SHISO and LenMa are both online clustering-based methods. For each incoming log message, SHISO \cite{6649746} and LenMa \cite{shima2016length} calculate a text or vector similarity against all the existing clusters of events and put the new log message into the cluster that has the highest similarity. If a significantly similar cluster is not found for a particular log message, both Shiso and Lenma formulate a new cluster representing its system event. LenMa uses a log message vectorization technique by encoding the number of characters in the words of a log message to a vector\cite{shima2016length}. It further uses the length heuristic to guarantee that the dimensions of the vectors to be sent to the vector similarity calculation are the same. Unlike LogSig, hierarchical clustering based methods such as LogMine and online clustering methods such as LenMa have shown to be more effective in practice~\cite{8804456}. The common challenge of clustering methods is identifying the terminating condition for the clustering algorithm. 

\subsection{Heuristics-based Techniques}

Some log parsing methods use heuristics in an algorithmic process to parse logs. Iteratively partitioning the log data set into event groups using heuristics is a commonly applied method. Makanju et. al. \cite{10.1145/1557019.1557154} used heuristics based on the message length and word position to iteratively partition the log data. He et. al. \cite{8067504} used a word count and word position heuristics in a similar way to parse logs. He et. al. \cite{8029742} used the message length and word position based heuristics as the basis for their method of using a fixed depth tree data structure for log parsing. Zhu et. al. \cite{8804456} showed that most of the state-of-the-art log parsing methods that rely on heuristics have significantly higher accuracy than the rest of the methods. But some heuristics created based on the assumptions such as "first couple of words of a log message is a \texttt{constant part}" or "the log messages of the same event have the same number of words", can fail. Since the log data is highly diverse and changing frequently, using heuristics for log parsing is not a sustainable option.
\subsection{Deep Learning-based Techniques}

Deep Learning (DL) techniques have also been utilized to solve the log parsing problem \cite{7916497, Thaler2017}. Thaler et. al. \cite{7916497}, has modeled the log parsing problem as a supervised sequential classification problem using a character-level Long Short-term Memory (LSTM) classifier. Once trained, the classifier can identify whether the characters present in a log message belong to a \texttt{constant} or \texttt{variable} token. Thaler et. al. \cite{Thaler2017} used an LSTM auto-encoder to generate semantic vector representation of log lines, after which the vector space is clustered and the event groups are extracted. This method is not a complete log parsing solution since it does not generate templates for the event groups. The lack of substantial amounts of labelled data for training leads to the accuracy of supervised DL methods to degrade. Furthermore, supervised deep learning techniques are not that practical for a form of data like log data that frequently changes with time where the addition of new logging events is more frequent. DL methods may also face challenges such as higher learning time and computation costs compared to  the other aforementioned methods.

\subsection{Information Retrieval Techniques}

The Paddy system~\cite{9110435} is the first to employ Information Retrieval techniques to address the system log parsing problem. It uses a dynamic inverted index data structure to store all the tokens present in the log message text in the form of terms. However the tokens containing wildcard characters are not indexed. The identified event IDs for an indexed term are stored in the posting list of that term. A new log message is first pre-processed and searched in the dynamic dictionary. Then the possible candidate event IDs are retrieved and the similarity between the input log message and each candidate template is calculated using a custom similarity metric called fitting score. Fitting score is the weighted sum of Jaccard similarity and length feature \cite{9110435}. If the highest similarity among the similarity values of all of the candidates is not greater than an experimentally decided threshold, the input log message is considered as a new template and gets indexed into the dynamic dictionary. Otherwise, the input log message is assigned to the correspondent event template. Then the words in the input log message and its template are checked by searching from left to right to identify the remaining \texttt{variable parts}. If the two words at the considered position of input log message and the event template are not the same, that word position of the event template is replaced by a wildcard to mark that as a \texttt{variable part}. This method has the highest recorded benchmark for log parsing when compared to the previous methods. However, Jaccard similarity is sub-optimal because it disregards term weighting. Furthermore, a log message and its identified template should have the same length to perform left to right check when identifying the remaining \texttt{variable parts}, by checking word positions of log message and identified event template from left to right.

In summary, automated log data structuring methods are introduced to avoid the challenges faced when using rule-based methods. The earliest automated log parsing methods used frequent pattern mining techniques. However, frequent pattern mining techniques fail to extract the event template of very rare events correctly. A major challenge to be faced when using clustering methods is identifying the terminating condition of the clustering algorithm. Heuristic-based methods are shown to be more accurate and robust, while they tend to fail on data that does not comply with the assumptions made in the heuristics. Deep learning methods have not been commonly used for log parsing due to the trade-off between computation cost versus accuracy gains. Recently introduced Information Retrieval techniques have shown to be the best.

\section{Vector Space Model for Document Similarity Measurement}
\label{theory}

The vector space model is a fundamental technique used to measure the similarity between a user query and a set of documents in Information Retrieval. First, both the query and the document are converted into vector representations. Then the vector distance between the query and each of the documents is measured using a vector distance measurement technique such as Euclidean distance or Cosine similarity. Then the document with the lowest vector distance to the query is selected as the candidate. The vector space model has also been employed in document similarity measurement  and sentence similarity measurement as well. The following discussion is presented with respect to document similarity measurement, for the sake of clarity.

\subsection{Deriving a Vector Representation}

TF-IDF (Term Frequency - Inverse Document Frequency) is a Bag-of-Word text representation technique that is widely used in the Information Retrieval and Natural Language Processing \cite{10.5555/866280, 8109100} to vectorize documents to derive a meaningful representation of text documents. 

In a bag-of-words model, the documents are represented by modeling a bag (un-ordered collection) of words. This model does not count the positioning, grammar or structure of the words in the text. It counts the frequencies of words in the target text and puts those words into a bag. The frequencies of words appearing in a text (sentence or document) is the feature that is used in the bag-of-words model.\\
As shown in equation \eqref{eq:tfidf}, a TF-IDF vector is calculated for a document as the product of Term Frequency (TF) and Inverse Document Frequency (IDF). According to the equation \eqref{eq:tf}, TF indicates the frequency of terms (i.e. how many times a term appeared in the document). Document frequency (DF) is the number of documents that contain a particular term (equation \eqref{eq:idf}). IDF is a measure of the rarity of the document terms with respect to the whole document collection.

\[ TF-IDF(t, d, D) = TF(t, d) * IDF(t,D) \tag{1} \label{eq:tfidf} \]

where,

\[ TF(t,d) = \frac{f_{t,d}}{\sum_{t'\epsilon  d}^{}f_{t',d\\}}  \tag{2} \label{eq:tf}\]

\[ IDF(t,D) = log(\frac{\left | D \right |}{df(t))}) + 1 \tag{3} \label{eq:idf}\]

$f_{t,d}$ is the number of occurrences of the term \texttt{t} in the document \texttt{d} which is one element of the target document set \texttt{D}. TF is a measure of how many times the term \texttt{t} appears in the document \texttt{d}. This can be used to identify the frequent and rare words present in a document. However, some stop words (e.g.~"the", "is") and some context-specific words may frequently appear in a document while adding no considerable information to that document.

IDF weighting can resolve this problem by emphasizing the informativeness and importance of words present in a document. 
IDF is defined for each term \texttt{t} in the vocabulary created for the set of documents \texttt{D}. IDF weighs a term \texttt{t} with respect to \texttt{D} so that the importance of \texttt{t} to a document \texttt{d} is encoded into the generated TF-IDF vector representation. If the term \texttt{t} is appearing in a significantly higher number of documents in \texttt{D}, \texttt{t} is most likely to be a  stop word or a context-specific word. According to the IDF equation \eqref{eq:idf}, the IDF value of that kind of word is smaller. A term that is unique to a document has a higher IDF value. 

By taking the product of the Term Frequency and Inverted Document Frequency terms, TF-IDF vector representation for a document \texttt{d} is formed. 

\subsection{Measuring distance Between Vectors}
As mentioned earlier, there are different types of vector distance measurement techniques. However, here we only discuss Cosine similarity, since it is the technique used in our system.

Cosine similarity is a widely used vector similarity measurement technique. It measures the closeness of two vectors by taking the normalized dot product (equation \eqref{eq:cs}). Cosine similarity value between two vectors lies between 0 and 1 inclusively. If the two vectors are highly similar to each other, the cosine similarity will be higher.

\begin{equation*} \tag{4} \label{eq:cs}
\begin{split}
CosineSimilarity (v1, v2) & = \frac{v1\cdot v2}{\left \| v1 \right \|\left \| v2 \right \|}  \\
 & = \frac{\sum_{i=0}^{n}v1_iv 2_i}{\sqrt{\sum_{i=0}^{n}v1^2}\sqrt{\sum_{i=0}^{n}v1^2}}
\end{split}
\end{equation*}

\section{Methodology}

\begin{figure}[!t]
\centering
\includegraphics[width=3.5in]{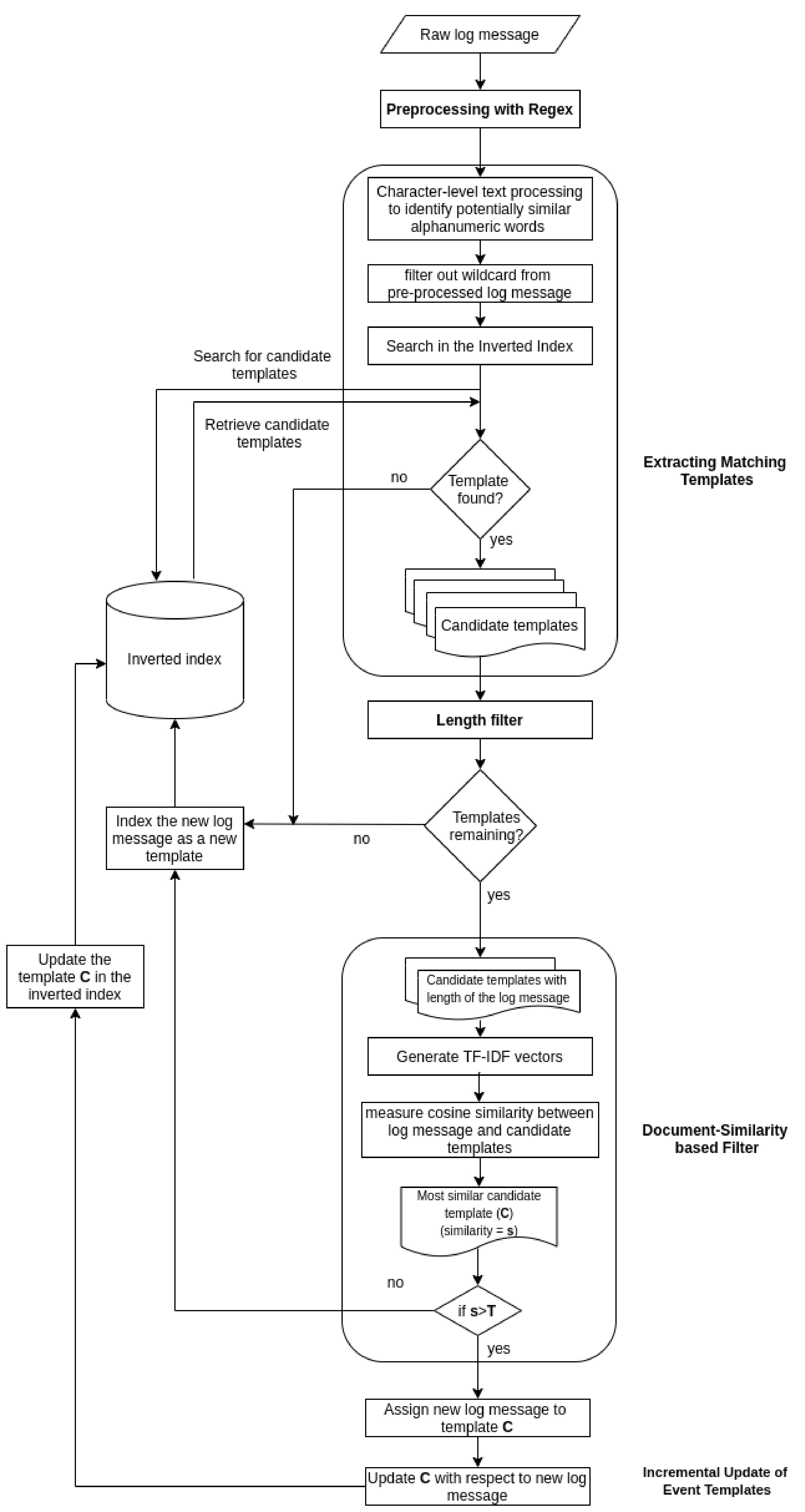}
\caption{vue4logs - Log parsing method}
\label{method}
\end{figure}

The log parsing technique employed in vue4logs makes use of Information Retrieval techniques to solve the log data structuring problem. This technique is similar to the log data structuring method proposed by Huang et. al. \cite{9110435}. However, they used Jaccard similarity based similarity measurement between a log message and the templates. Jaccard similarity does not identify the importance of words to a given text. To overcome these shortcomings, the vector space model for document similarity measurement is presented in this paper. Furthermore vue4logs uses a character-level text processing technique to have a better parsing accuracy. Vue4logs uses inverted index data structure by following Huang et. al. \cite{9110435}.
 
vue4logs comprises five main steps as shown in Fig. \ref{method}. First, the new log message is pre-processed using regular expressions used by Zhu et. al. \cite{8804456}. Secondly, the log message is searched in the inverted index and matching candidate event templates are retrieved. If there is no matching event template retrieved in the search, the log message itself is indexed as a new template in the dynamic inverted index. In the third step, the candidate event templates that do not have the same length as the log message are filtered out using the length filter. Next, TF-IDF technique is used to vectorize the pre-processed log message and the length-filtered candidate templates. Then the correct event template out of the candidates is identified by measuring the Cosine similarity between the candidate templates and the log message. If the log message is not similar enough to any candidate template, it is considered as a new event template and gets indexed in the dynamic inverted index. In the final step, the selected template is updated with respect to the newly assigned log message. Then the dynamic inverted index is updated with respect to the update done to the template.

\subsection{Pre-possessing using Regular Expressions}

Log data is pre-processed using regular expressions to identify potential \texttt{variable parts} such as IP addresses. To perform a fair evaluation, the same regular expression set used by Zhu et. al. \cite{8804456} and Huang et. al. \cite{9110435} is used when evaluating vue4logs. Zhu et. al. \cite{8804456} has used a minimal set of regular expressions for pre-possessing to have fairness among the parsers. Once the raw log message is pre-processed, it becomes a partial event template, which will be further processed in the next steps.

\subsection{Extracting Matching Templates}
\label{ext-mat-temp}

An inverted index is a data structure commonly used in the Information Retrieval domain. Inverted index is used to store the mapping from a term to the documents that term appears in. In our use-case, the \texttt{constant parts} present in the identified event templates act as the terms in the inverted index data structure. A unique ID is used to identify an event template. The ID of an event template that has a particular term inside it is put into the posting list of the particular term. Inverted index is used to retrieve the documents (identified event templates) that have a particular term inside it. For example, if the identified templates are, \texttt{Connection broken from user id <*> myid <*> error} and \texttt{Invalid user <*> from <*>}, Table \ref{invereted_index} is the inverted index created for those two templates.

\begin{table}[t!]
\centering
\begin{tabular}{ | p{6em} | p{1.7cm}| p{1.7cm} | }
\hline
Term	& Posting List \\
    \hline
    Connection	 & [1]	\\
    broken	 & [1]	\\
    from	& [1,2]	\\
    user	 & [1,2]	\\
    id	 & [1]	\\
    myid	 & [1]	\\
    error	 & [1]	\\
    Invalid	 & [2]	\\
    \hline

 \end{tabular}
 \caption{Inverted Index for the Sample Event Templates Described in the Section 4.2}
 \label{invereted_index}
 \end{table}
 
Initially, the log message is tokenized, and the numeric characters present in the alphanumeric tokens are replaced with the wildcard character. As done in \cite{7916497}, the consecutive wildcards found in character-level check are merged together to form a representative term for a particular token. For instance, if the pre-processed log message is \texttt{updateNotificationShade: total=1, active=1}, the two tokens \texttt{total=1,} and \texttt{active=1} are converted to \texttt{total=<*>,} and \texttt{active=<*>} respectively by using the above process. Once this character-level processing technique is performed, the term \texttt{total=<*>,} can be used to  represent the tokens such as \texttt{total=0,} and \texttt{total=1,} while the term \texttt{active=<*>} can be used to represent the tokens such as \texttt{active=0} and \texttt{active=1}. This character-level text processing technique helps to handle the slight variations present in nearly identical words.

Next, the resultant log message is passed through a wildcard filter that filters out the wildcard tokens from the log message. This ensures that no template is retrieved because of matching wildcards in the pre-processed log message and a template. The output of the wildcard filter is used as the search query to search the inverted index. For instance, if the log message before going through the wildcard filter is \texttt{Connection broken for id <*> my id = <*> error}, the search query after wildcard filtering is \texttt{Connection broken for id my id = error}. 

Once the search is performed, the event templates that have at least one term matching with the search query are retrieved as the candidate event templates for a particular log message. The retrieved candidate event templates are sent to the next stage along with the log message. 

If no event template is retrieved as a candidate, the new log message is considered as the first occurrence of its event after-which, a new template ID is generated and indexed into the inverted index. Then, the new template is passed through the wildcard filter and only the \texttt{constant} tokens that need to be indexed are identified.  When indexing, each \texttt{constant} token is checked with the terms present in the inverted index. If the token is already present in the inverted index as a term, the ID of the new template is appended to the posting list of that term. Otherwise, the token is added to the inverted index as a new term with a posting list containing only the newly generated event template ID.

\subsection{Length Filter}

If there are one or more candidate templates retrieved in the search step, whether the correct event template is present among the candidate event templates should be identified. To do this, the retrieved candidate templates are passed through a length filter. It is generally observed that the log messages of the same event have the same length. Length in this context is the number of words present in the log message. This property is vastly used to identify the event groups of log messages \cite{10.1145/1557019.1557154, 8029742}. vue4logs also uses this property as a length filter to filter out the candidate templates with different lengths than the log message. This step achieves two objectives. Firstly, it narrows down the search space of possible candidate event templates before moving to the log data vectorization step, which is a relatively computationally expensive process. Secondly, it ensures that the selected event template and the log message have the same length so that when updating the templates with respect to the newly assigned log message by checking words from left to right, the words can be identified by their position.

After the length filter, the resultant candidate event templates are greedily matched with the log message to decide whether any of the length-filtered event templates are textually identical to the log message. If so, the log message is assigned to that event template. This is done because it is more efficient to identify the correct event template by textual matching than doing text vectorization and vector similarity measurements. If none of the length-filtered templates is identical to the log message textually, candidate event templates are sent to log vectorization technique along with the log message. 

If no candidate event template is remaining after the length filtering, the log message is considered as the first occurrence of its event and is indexed as a new template as discussed in the previous section.

\subsection{Document-Similarity based Filter}
Next, the log message and the candidate event templates are tokenized using the space character. The uppercase and lowercase words such as \texttt{Connection} and \texttt{connection} are considered as two different tokens in the created vocabulary because, unless the word \texttt{Connection} (or \texttt{connection}) is a \texttt{variable part}, all the log messages of the same event have either an uppercase word or lowercase word at a specific position but not the both. The meaning of the text is not the key concern of this use case since the goal is to identify the structure of log messages. 

The set of all the unique tokens after tokenization is included in the  vocabulary. The wildcard token replaces the numeric characters in alphanumeric tokens, and consecutive wildcards inside the token are merged together as described earlier. Then, the TF-IDF vectorization process discussed in Section \ref{theory} is performed to generate vectors for the log message and candidate event templates.   It has been observed that frequent words in a log data set are more likely to be \texttt{constant parts} and rare words are more likely to be \texttt{variable parts} \cite{1251233}. The ability of TF-IDF to encode the importance of words can exploit this property of log messages. Therefore, TF-IDF is better at representing log messages and event templates compared to using lexical similarity. The log message and the set of length filtered candidate event templates are the documents to be sent for log data vectorization. The dimension of the vector space depends on the number of tokens present in vocabulary. 

Once the log data is encoded into the TF-IDF vector representation, the vector similarity between the vector of the log message and the vector of each candidate event template is calculated. Cosine similarity is the vector similarity technique used in this method.

If the highest cosine similarity value among the calculated values is greater than an empirically decided threshold \textbf{T}, the new log message is assigned to the corresponding event template. Once the correct event template is identified, it is updated with respect to the new log message (as discussed in the next section). If the highest similarity is less than \textbf{T}, the new log message is considered as the first occurrence of its event and gets indexed into the inverted index as a new event template.

\subsection{Incremental Update of Event Templates}

In most of the online log parsing methods, the identified event template is incrementally updated with respect to the newly assigned log messages \cite{shima2016length, 9110435} until the correct version of the template is formed. In vue4logs, this process operates by checking the words of the log message and the event template left to right by assuming that the lengths of log messages of the same event are the same. The length filter used in the earlier step guarantees the length equality of the log message and the selected template so that this step can be performed. This guarantees that the words present in the log message and the selected event template can be identified by the position of the words. This allows vue4logs system to identify the remaining \texttt{variable parts} by checking the log message and the template from left to right.

For instance, if the current log message is \texttt{Invalid user webmaster from <*>} and the selected  event template is \texttt{Invalid user chen from <*>}, the words \texttt{webmaster} and \texttt{chen} are considered to be at the second position of the two documents. When updating the event template, each word pair at the same position is compared. If the two words are the same, the selected event template is not updated for that word position. For instance, the word \texttt{user} (at position one) remains unchanged in the event template. If the words in the concerned position differ from each other, the event template is updated by replacing the word in that position with the \texttt{<*>}. For instance, as words in the position two are not the same (\texttt{webmaster} and \texttt{chen}), the selected event template is
updated by replacing the term \texttt{chen} with the wildcard (\texttt{<*>}). The resultant event template is \texttt{Invalid user <*> from <*>}. The word \texttt{chen} in the original event template is now identified as a \texttt{variable part} and therefore, the template ID of the selected event template is removed from the posting list of the term \texttt{chen} in the dynamic inverted index. By doing this, it is ensured that the selected template is no longer retrieved in the search step, just because the term \texttt{chen} appears in a new log message. Once the log parsing method has seen enough log messages of a particular event, it eventually generates the final event template.

\section{Experiments and Results}

\subsection{Experiment Setup}

\begin{table*}[t!]
\centering
\begin{tabular}{ |c|c|c|c|c|c|c|  }
\hline
data set &Description& Time Span &Data size & \#Messages & \#Templates (total) & \#Templates (2k) \\
\hline
\multicolumn{7}{|c|}{Distributed system logs} \\
    \hline
    HDFS & Hadoop distributed file system log &38.7 hours &1.47 GB& 11,175,629& 30& 14 \\
    \hline
    Hadoop & Hadoop mapreduce job log &N.A. & 48.61 MB & 394,308 &298 &114 \\
    \hline
    Spark & Spark job log &N.A. &2.75 GB &33,236,604 &456 &36 \\
    \hline
    ZooKeeper & ZooKeeper service log &26.7 days &9.95 MB& 74,380 &95& 50 \\
    \hline
    OpenStack & OpenStack software log & N.A. & 60.01 MB & 207,820 & 51 & 43 \\
 \hline
 \multicolumn{7}{|c|}{Super computer logs} \\
    \hline
    BGL & Blue Gene/L supercomputer log &214.7 days &708.76 MB &4,747,963 &619 &120 \\
    \hline
    HPC & High performance cluster log &N.A.& 32.00 MB &433,489 &104& 46 \\
    \hline
    Thunderbird &Thunderbird supercomputer log& 244 days &29.60 GB& 211,212,192 &4,040& 149 \\
  \hline
  \multicolumn{7}{|c|}{Operating system logs} \\
    \hline
    Windows & Windows event log & 226.7 days & 26.09 GB & 114,608,388 & 4,833 & 50 \\
    \hline
    Linux & Linux system log & 263.9 days & 2.25 MB & 25,567 & 488& 118 \\
    \hline
    Mac& Mac OS log &7.0 days &16.09 MB& 117,283 &2,214 &341 \\
  \hline
  \multicolumn{7}{|c|}{Mobile system logs} \\
    \hline
    Android & Android framework log &N.A. & 3.38 GB &30,348,042& 76,923& 166 \\
    \hline
    HealthApp &Health app log &10.5 days& 22.44 MB& 253,395 &220& 75 \\
  \hline
  \multicolumn{7}{|c|}{Server application Logs} \\
    \hline
    Apache & Apache server error log & 263.9 days& 4.90 MB& 56,481 &44 &6 \\
    \hline
    OpenSSH &OpenSSH server log &28.4 days &70.02 MB& 655,146 &62 &27\\
  \hline
  \multicolumn{7}{|c|}{Standalone software Logs} \\
    \hline
    Proxifier &Proxifier software log& N.A. &2.42 MB &21,329& 9 &8\\
    \hline
 
 \end{tabular}
 \caption{Characteristics of Data According to \cite{8804456}}
 \label{data}
 \end{table*}
 
\subsubsection{Implementation Details}

Following previous studies \cite{8804456, 9110435}, the parameters of all the log parsing methods used in this evaluation are fine-tuned through over 10 runs per data set and the best results were reported to avoid bias from randomization. The similarity threshold, the parameter to be tuned in vue4logs was decided using the same process. The scikit-learn implementation \cite{sklearn_api} \footnote{scikit-learn version 0.23.2} is used to generate the TF-IDF vector representations. All the experiments were conducted on a server with 8 Intel(R) Xeon(R) 2.30GHz CPUs, 32GB RAM, and Ubuntu 18.04.3 LTS installed.

\subsubsection{Baseline}

vue4logs is evaluated against two state-of-the-art methods, namely Drain \cite{8029742} and Paddy \cite{9110435}, which are the methods with overall best results according to the literature \cite{8804456, 9110435}. Since the evaluation procedure proposed by Zhu. et. al. \cite{8804456} and Huang et. al. \cite{9110435} is followed, the same set of regular expressions used in their evaluation was used in the pre-possessing step of vue4logs. This is because the regular expressions used in the reprocessing steps have a significant impact on the accuracy of the log parsing method. 

\subsubsection{Data}

The log data used in this evaluation comprises the data extracted from real-world systems. The dataset was published by Zhu et. al. \cite{8804456} and can be found in \footnote{https://github.com/logpai/loghub}. This data set comprises log data extracted from sixteen different systems including distributed systems, supercomputers, operating systems, mobile systems and standalone software. The variety of this data set is a useful feature to measure the ability of a solution to extract event templates from heterogeneous log data. The complete log data set covers 440 million loglines. Table \ref{data} shows a summary of the characteristics of data. Zhu et. al \cite{8804456} have randomly sampled 2000 log messages from each data set and manually labelled the correspondent event templates in order to derive the ground truth. The "\#Templates (total)" column of Table \ref{data} shows the total number of event templates generated by a rule-based log parser, and "\#Templates (2k)" column shows the number of event templates present in the log data sample of size 2000. The evaluation of Drain \cite{8804456} and Paddy \cite{9110435} is also done on this data set and therefore the accuracy benchmarks of those methods are already available. 

In addition to this commonly used benchmark, in order to experiment with the robustness of vue4logs, a new log data set which was introduced in \cite{Thaler2017} is also used. 

\subsubsection{Evaluation Metrics}

The vue4logs system is evaluated with respect to three aspects: parsing accuracy, robustness, and efficiency. In general, the parsing accuracy is the measure of identifying the event templates with the exact number of log messages that should belong to that template. Accuracy is a critical evaluation aspect as it affects the usability of structured log data output on subsequent log analysis use cases \cite{7579781}. 

F1-Score is used in several studies to evaluate the accuracy of the log parser \cite{8804456}. However this is a measure of collective accuracy of clustering output for individual log messages. Having a higher F1-score does not necessarily mean that an extracted event template can be used to represent all the log messages of the target event group. Therefore, Paring Accuracy (PA) has been employed by some previous research~\cite{8489912,8804456,9110435}.

\[ PA =\frac{Number \;  of \; correctly \; parsed \; log \; messages}{ Total \; number \; of \; log \; messages} \]

Parsing Accuracy considers an extracted template correct if and only if all the log messages corresponding to the event template are included under the extracted template \cite{8489912, 8804456}. An assignment of a single incorrect log message to a template or any event template that is unable to cover all the log messages as ground truth states leads the whole event template to be inaccurate. For instance, if the log message sequence \texttt{[ m0, m1, m2, m3, m4, m5 ]} is mapped to events  \texttt{[E0, E1, E1, E1, E1, E2 ]} while correspondent true event sequence is \texttt{[E0, E1, E1, E1, E2, E2 ]} the parsing accuracy for this grouping is $1/6$ since only the event \texttt{E0} is correctly identified without assigning more or less number of log messages to the event template compared to the truth. \texttt{m4}, which should have been put into the template of \texttt{E2} is now put into the template of \texttt{E1}, making both \texttt{E1} and \texttt{E2} event templates incorrect. Because of this property, Du et. al. \cite{8489912} and Zhu et. al. \cite{8804456} state that this metric is a more rigorous one to be used to evaluate the accuracy of the log parsing. Therefore the evaluation of the vue4logs is also done using the PA as the evaluation metric.

Robustness of a log parsing method measures the consistency of its accuracy under log data extracted from different systems. A robust log parsing method should perform consistently across different log data from different sources. 

Efficiency measures the processing speed of a log parsing method. Efficiency is evaluated by recording the time that a log parsing method takes to parse a log file of a specific data set. The less time a log parsing method consumes, the higher the efficiency it provides.

\subsection{Experimental Results}

\begin{table}[t!]
\centering
\begin{tabular}{ |l||l|l|l|l| }
\hline
Data set & Drain & Paddy & all-time best & vue4logs\\
    
    \hline
    HDFS &	\textbf{0.998} &	0.94 & 1	& \textbf{0.998} \\
    \hline
    Hadoop &	0.948 &	0.952 &	0.957	& \textbf{0.998} \\
    \hline
    Spark &	0.92 &	0.98 &	0.994	&\textbf{0.998} \\
    \hline
    Zookeeper &	0.967 &	0.986 &	0.986	&\textbf{0.993} \\
    \hline
    OpenStack &	0.733 &	0.839 &	0.871	&\textbf{0.958} \\
    \hline
    BGL &	0.963 &	0.963 &	0.963	&\textbf{0.986} \\
    \hline
    HPC &	0.887 &	0.904 &	0.904	&\textbf{0.932} \\
    \hline
    Thunderbird  &	0.955 &	0.951 &	0.955	&\textbf{0.957} \\
    \hline
    Windows	 & 0.997 &	0.994 &	0.997	&\textbf{1} \\
    \hline
    Linux	 & 0.69 &	\textbf{0.859} & 0.859	&0.675 \\
    \hline
    Mac	 & 0.787 &	0.851 &	0.872	&\textbf{0.928} \\
    \hline
    Android	 & 0.911 &	0.878 &	0.919	&\textbf{0.93} \\
    \hline
    HealthApp	 & 0.78 &	0.755 &	0.822	&\textbf{0.99} \\
    \hline
    Apache	 & \textbf{1}  &	\textbf{1}  &	1	&\textbf{1} \\
    \hline
    OpenSSH	 & 0.788  &	\textbf{0.938} & 0.938		&0.925 \\
    \hline
    Proxifier &	\textbf{0.527}  &	\textbf{0.527}  & 0.967		&\textbf{0.527} \\
    \hline
    \hline
    Average &	0.865 &	0.895 &	N.A	& \textbf{0.924} \\
 \hline
 
 \end{tabular}
 \caption{ACCURACY OF LOG PARSING METHODS ON DIFFERENT DATA SETS}
 \label{pa_ss}
 \end{table}

\subsubsection{Accuracy}

Table \ref{pa_ss} shows the parsing accuracy results of vue4logs and the selected baseline systems. The `all-time best' column reports the highest result ever reported for each dataset. In other words, this is the maximum reported result of the empirical study by Zhu et. al.~\cite{8804456} and the later published Paddy system.  It can be observed that vue4logs significantly outperforms the baseline methods with an average accuracy of 0.924, while the next highest is the Paddy with an average accuracy of 0.895. Moreover, vue4logs sets a new baseline result for 12 of the datasets (counting the result for Apache dataset as well).

vue4logs is able to parse Apache and Windows log data sets with a 100\% accuracy. Apache is a relatively simple log data set which is parsed with 100\% accuracy by most of the methods according to the benchmarks \cite{8804456, 9110435}. However, none of the previously proposed methods has achieved 100\% parsing accuracy on the Windows dataset.

Datasets such as Mac and HealthApp have obtained more than a 5\% accuracy improvement when compared with the all-time best parsing accuracy on those two datasets. Manual observations showed that HealthApp and Mac data have a significantly higher number of words that have varying numeric parts inside them. The higher accuracy improvement on HealthApp and Mac is mostly due to the use of character-level numerical check and merging done before creating the vocabulary. 

vue4logs has the highest all-time parsing accuracy for eight other data sets namely Hadoop, Spark, Zookeeper, OpenStack, BGL, HPC, Android and Thunderbird. The main reason for this accuracy improvement is that vue4logs used vector space based similarity measurement techniques instead of Jaccard coefficient. TF-IDF vectorization is able to encode the frequency and importance of terms, while Jaccard simply relies on text overlap. Thus, the difference between \texttt{constant} and \texttt{variable} parts is better encoded with TF-IDF. Furthermore, as mentioned above the improvement done by using character-level techniques when creating the vocabulary has significantly influenced to improve the effectiveness of cosine similarity on the generated vectors.

Parsing accuracy benchmarks of HDFS (99.8\%)  and OpenSSH (92.5\%) are at an acceptable level compared to the all-time highest benchmarks (100\% and 93.8\% respectively).

The parsing accuracy benchmarks of vue4logs on the Proxifier and Linux data sets are not higher than that of the best benchmarks published earlier. When analysed the true templates of Proxifier data set, it can be observed that it has a small number of events (8) in 2000 log lines while half of them are textually very similar to each other. A similar observation was made for the Linux data set. Furthermore, Linux data has a complex structure of logging event patterns, which is relatively difficult to identify without dataset specific features. 

\subsubsection{Significance Test}
\label{t-test}
To decide whether vue4logs is significantly accurate than the best baseline - Paddy, a t-test was performed on the parsing accuracy results of Paddy and vue4logs. The data points in the two samples are paired based on the data set. Furthermore, the objective is to check whether vue4logs  performs better than Paddy. Therefore, a one-tailed, paired t-test was carried out on the two parsing accuracy benchmarks. The null hypothesis was that there is no significant improvement in log parsing accuracy of vue4logs when compared to that of the Paddy method. The t-test results showed that results of the proposed method can be random with a 0.09 probability. Therefore, it could be concluded that vue4logs has a significant improvement in log parsing accuracy over Paddy with 90\% confidence.

\subsubsection{Robustness}
 
\begin{figure}[!t]
\centering
\includegraphics[width=3.5in]{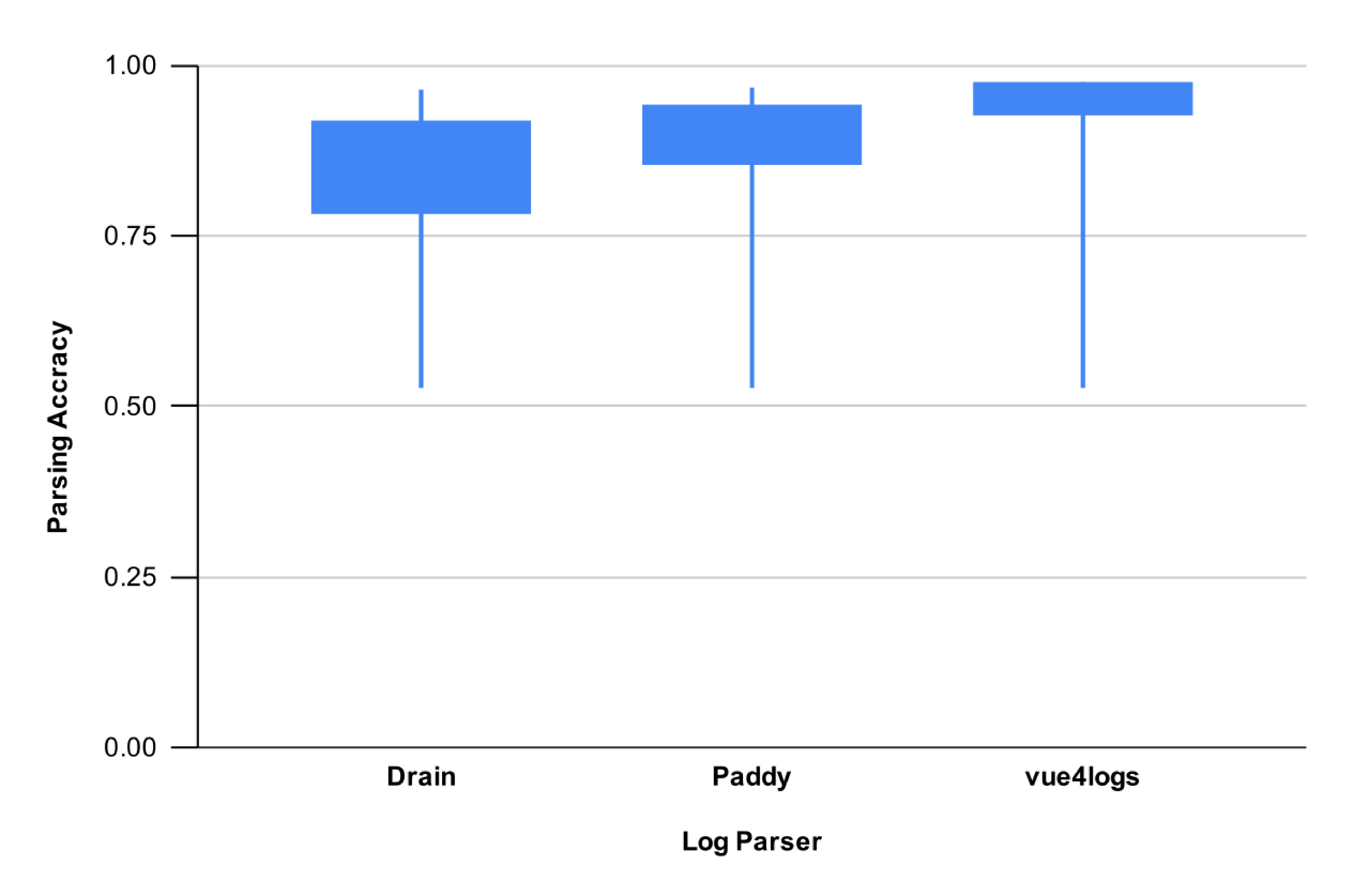}
\caption{ Accuracy Distribution of Evaluated Methods across Different Types of Logs}
\label{acc_dis}
\end{figure}

Robustness is a critical property of a log parsing method, which can be used as a suitability criteria to decide whether a given log parsing method can be used in a production environment. The
robustness of vue4logs is evaluated by measuring its ability to be used on heterogeneous log data.

One way of evaluating the robustness is to measure the variance of parsing accuracy distribution of a log parsing method over data sets from different sources. Since the selected data sets comprise a highly diverse set of log data from different types of systems, it gives an ideal environment for evaluating robustness of the log parsers.
The box-plot illustrated in Figure \ref{acc_dis} shows the distribution of accuracies of the evaluated methods across the sixteen data sets. From left to right in the Figure \ref{acc_dis}, the evaluated log parsing methods are arranged in the ascending order of the average accuracy shown in Table \ref{pa_ss}. That is, Drain has the lowest accuracy and vue4logs obtains the highest accuracy on average.  vue4logs also shows the smallest variance across the dataset, which is an indication of its usability on heterogeneous log data.

Another approach for evaluating the robustness of the log parsing methods is to identify data independent values for the configuration parameters of the log parsers and use those parameter values on an unseen custom dataset. 

\begin{table}[t!]
\centering
\begin{tabular}{ | p{6em} || p{1.7cm}| p{1.7cm} | p{1.7cm} | }
\hline
Data set	& Drain (depth = 4, = 0.81) &	Paddy (T= 0.79) & vue4logs  (T = 0.61) \\
    \hline
    HDFS	 & 	\textbf{0.998}	 & 	0.94	 & 	\textbf{0.998} \\
    \hline
    Hadoop	 & 	0.953	 & 	0.85	 & 	\textbf{0.959} \\
    \hline
    Spark	 & 	0.907	 & 	0.718	 & 	\textbf{0.92} \\
    \hline
    Zookeeper	 & 	0.967	 & 	\textbf{0.985}	 & 	0.967 \\
    \hline
    OpenStack	 & 	0.733	 & 	0.881	 & 	\textbf{0.903} \\
    \hline
    BGL	 & 	0.961	 & 	0.813	 & 	\textbf{0.985} \\
    \hline
    HPC	 & 	0.9	 & 	0.833	 & 	\textbf{0.906} \\
    \hline
    Thunderbird	 & 	0.953	 & 	0.652	 & 	\textbf{0.958} \\
    \hline
    Windows	 & 	0.569	 & 	\textbf{0.99}	 & 	0.693 \\
    \hline
    Linux	 & 	\textbf{0.69}	 & 	0.236	 & 	0.236 \\
    \hline
    Mac	 & 	0.693	 & 	0.856	 & 	\textbf{0.871} \\
    \hline
    Android	 & 	\textbf{0.905}	 & 	0.778	 & 	0.844 \\
    \hline
    HealthApp	 & 	0.485	 & 	0.175	 & 	\textbf{0.995} \\
    \hline
    Apache	 & 	\textbf{1}  &	\textbf{1}  &	\textbf{1} \\
    \hline
    OpenSSH	 & 	0.788	 & 	0.677	 & 	\textbf{0.925} \\
    \hline
    Proxifier	 & 	0.026	 & 	\textbf{0.049}	 & 	0.026 \\
    \hline
    \hline
    Average	 & 	0.783	 & 	0.714	 & 	\textbf{0.824} \\
    \hline

 \end{tabular}
 \caption{ACCURACY ON DIFFERENT DETESTS UNDER SOURCE INDEPENDENT PARAMETER TUNING}
 \label{pa_si}
 \end{table}

As the custom data set,  the“Unix forensic logs” dataset \cite{Thaler2017} was used. This data set has been extracted from an Ubuntu 16.04 system image disk. Thaler et. al. \cite{Thaler2017} have manually created the event IDs as cluster labels for this data set by looking at the Ubuntu source code. According to the authors, the difference between a forensic log and a system log is that a forensic log contains information from multiple log files on the examined system, whereas a system log only contains the logs that were reported by the system daemon. The system log is part of the forensic log, but it also contains other logs, which typically leads to more complexity in such log files. When the logs of this data set are observed, it can be seen that it has a large variety of logs generated inside a Unix based system. Because of this reason, this data set is more suitable to be used as the custom data set in the robustness experiments.

However, Thaler et. al. \cite{Thaler2017} only measured the clustering quality on log data and therefore, there are no event templates labelled in the dataset. For our evaluation, the dataset has to be labelled with true event templates as well. To form the ground truth for the log parsing problem, a set of log lines of size 2000 was randomly sampled. Originally, Thaler et. al. \cite{Thaler2017} have removed the date and timestamp headers from the log lines of this data set. By manually observing the logs of the data set, we removed the other specific header fields such as program and PID, verbosity level and duration from the log messages in the selected sample. Then, the event templates of the 2000 log messages were manually labelled by referring to the already labelled event IDs of the data set to form the ground truth. This annotated dataset is publicly released.

\begin{table}[t!]
\centering
\begin{tabular}{ | p{7em} | p{1.7cm}| p{1.7cm} | p{1.7cm} | }
\hline
Data set	& Drain \newline (depth = 4, \newline st = 0.81) &	Paddy  \newline (recreation) \newline (T= 0.79) &	vue4logs  \newline (T = 0.61) \\
    \hline
    Unix Forensic	 & 0.651	 & 0.679	 & 	0.713 \\
    \hline

 \end{tabular}
 \caption{ACCURACY ON THE CUSTOM DATA SET }
 \label{pa_hold_out}
 \end{table}
 
 \begin{figure*}[!t]
\centering
\includegraphics[width=2.3in]{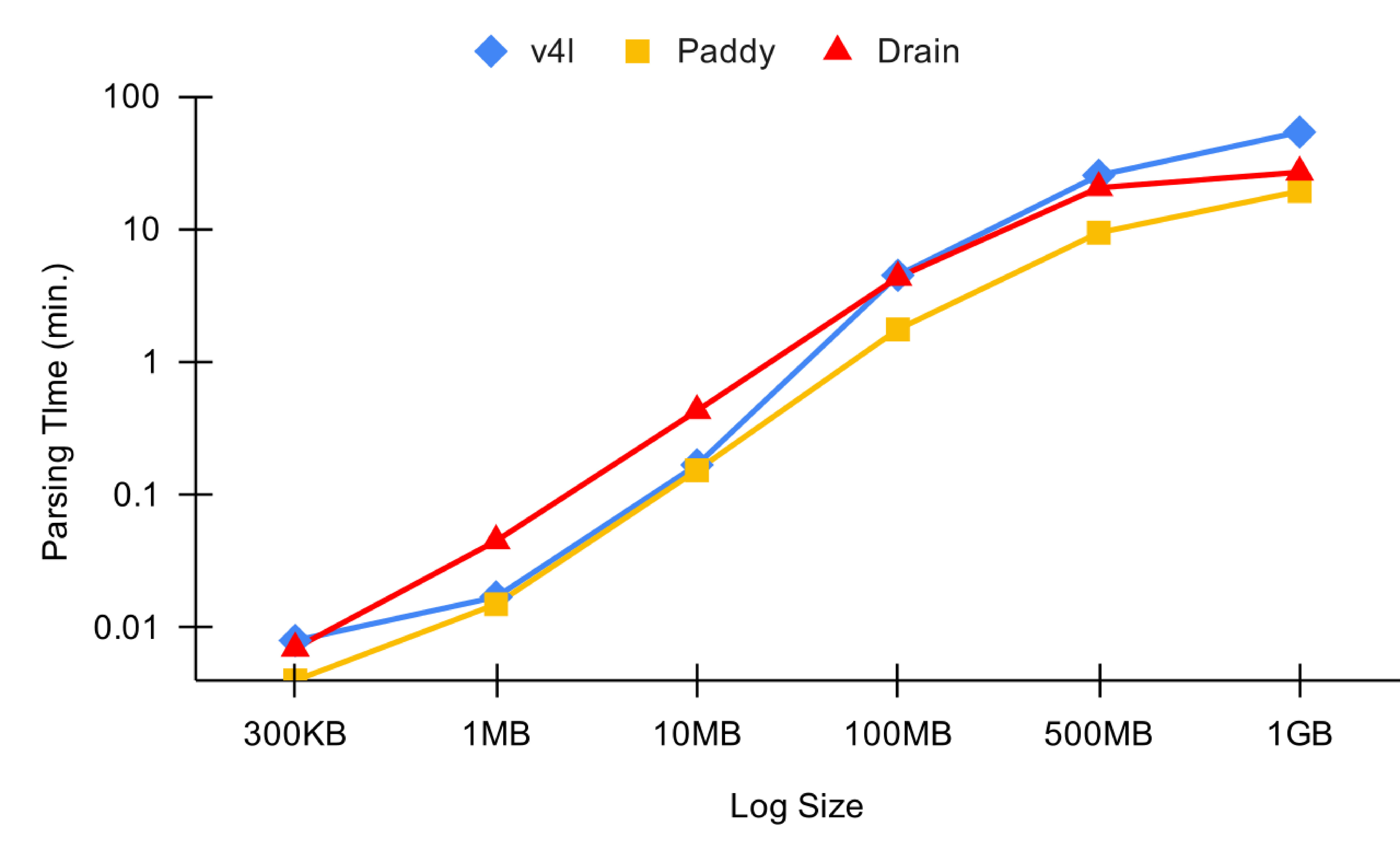}
\includegraphics[width=2.3in]{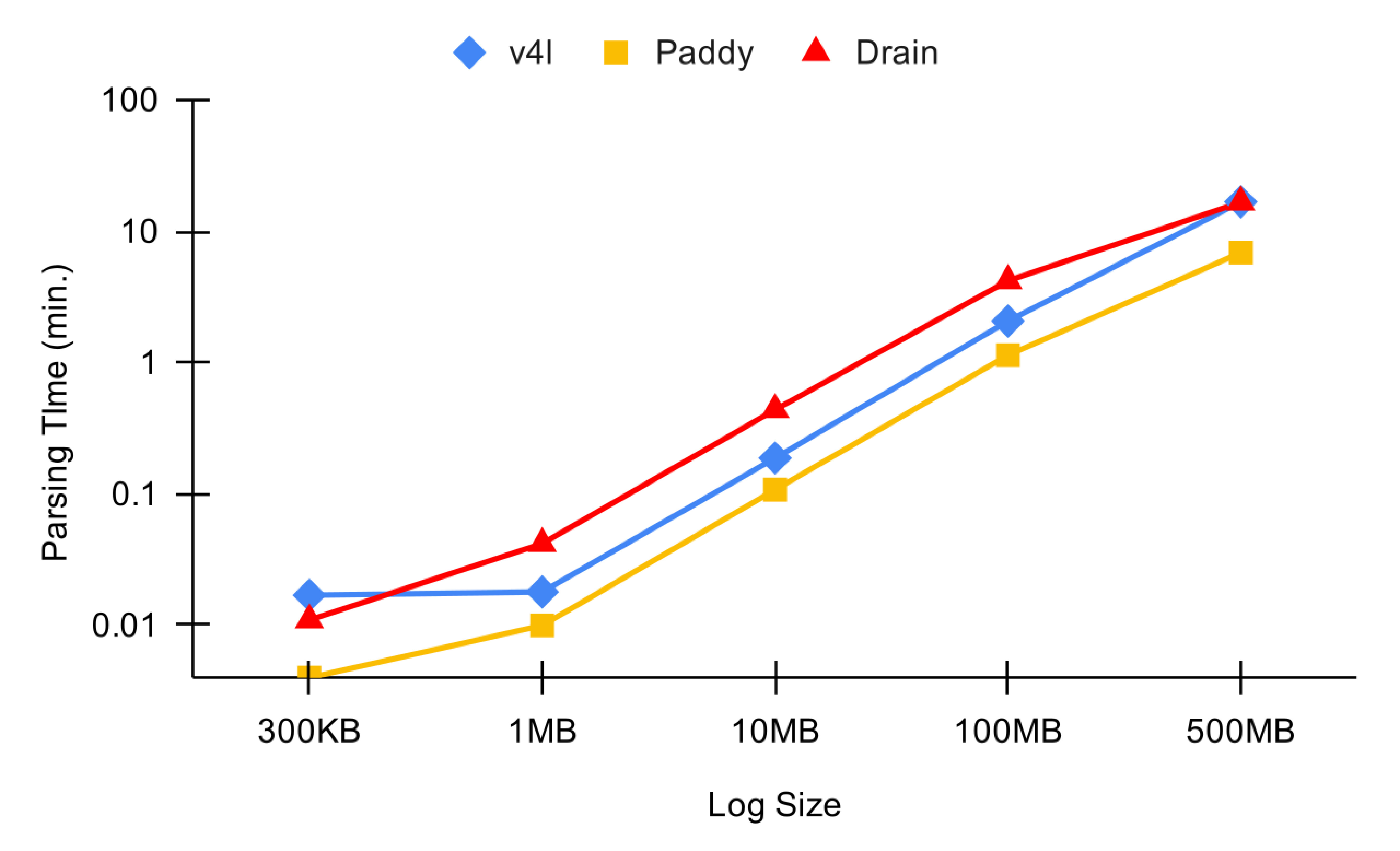}
\includegraphics[width=2.3in]{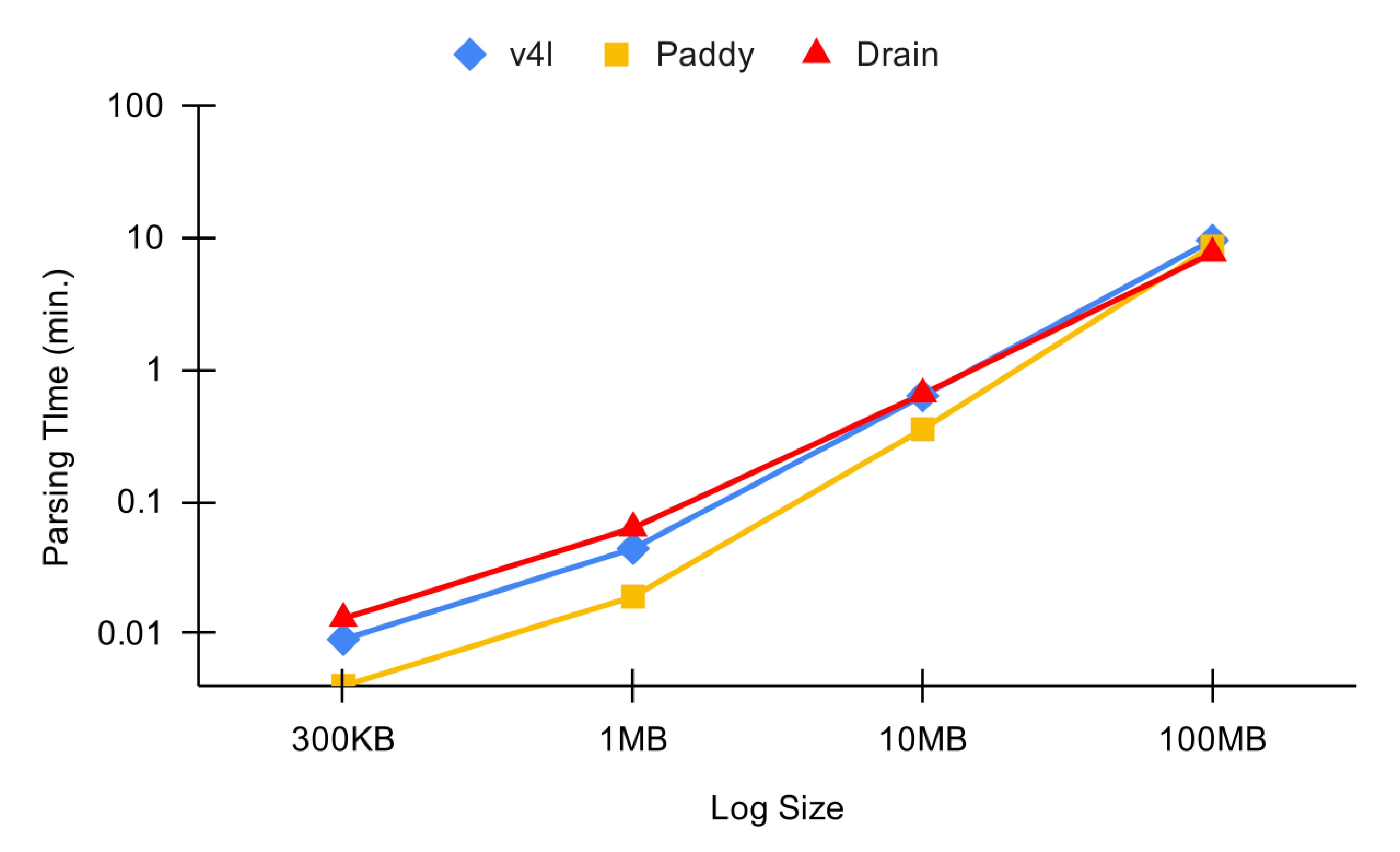}
\caption{ Accuracy Distribution of Evaluated Methods across Different Types of Logs}
\label{eff}
\end{figure*}

Drain log parsing method has two configurable parameters namely, tree depth and similarity threshold (st). In contrast, Paddy and vue4logs have only one tunable parameter which is the similarity threshold (T). Paddy method was recreated by following the method published in \cite{9110435}. 
The parsing accuracy (PA) benchmarks on the custom data set shown in the Table \ref{pa_hold_out} are lower compared to their source-independent parameter tuning benchmarks shown in Table \ref{pa_si}. This is mainly because this particular data set contains a more diverse set of log messages when compared to the other log datasets. However, out of evaluated methods, vue4logs has the best parsing accuracy on the custom data set. This means the generalizability of vue4logs on custom log data is higher than that of the other two methods evaluated. Therefore, the usability of vue4logs on a custom log data is shown to be higher than the state-of-the-art log parsing methods.
 
\subsubsection{Efficiency}
The efficiency of the three log parsers is measured on different sizes of HDFS, BGL and Android data sets as done in \cite{8804456} and \cite{9110435}. The results of the efficiency experiments are presented in Figure \ref{eff}. The processing time axis is presented in the logarithmic scale. It is obvious that the parsing time increases with the raising of log size on all three data sets. It can be observed that vue4logs has reasonable efficiency benchmarks when compared to Drain and Paddy. In particular, the efficiency benchmarks of vue4logs lie in between that of Drain and Paddy which means it has an acceptable level of efficiency. TF-IDF technique and character-level text processing techniques, which are relatively expensive computations bound the efficiency of vue4logs.

\section{Conclusion}

The log data structuring problem is a well-known problem in the log analysis domain. Even though a variety of methods and techniques have been utilized to address this problem, most of those methods use techniques fine-tuned on a given dataset.

The method proposed in this paper uses Information Retrieval techniques, more specifically vector space based similarity calculation using TF-IDF and cosine similarity for log parsing. It can be observed that the most general heuristics such as length heuristic can be used to simplify the log parsing process without significantly affecting the accuracy. Furthermore, it is observed that the words present in the log messages generated by the same events have slight character-level variations. A character-level text processing techniques is introduced to exploit this property in order to improve the vocabulary creation steps of vue4logs.

vue4logs has the highest results with respect to parsing accuracy. The results of the statistical t-test show that vue4logs is significantly accurate than the baseline method called Paddy. Having the lowest variance of accuracy across different log data sets, as well as having the highest results on a custom data set under source-independent parameter tuning setting show that vue4logs can handle custom data more consistently when compared to the considered log parsing methods.

Similar to some of the state-of-the-art log parsing methods, vue4logs also fails to generate correct event templates for the event groups that have log messages of different lengths. vue4logs is computationally expensive compared to some of the other methods. These two problems will be handled in the future.

\ifCLASSOPTIONcaptionsoff
  \newpage
\fi



%
\bibliographystyle{IEEEtran}
\bibliography{ref}

%




\begin{IEEEbiography}
[{\includegraphics[width=1in,height=1.25in,clip,keepaspectratio]{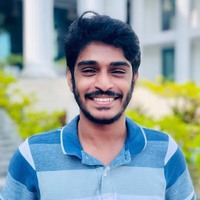}}]{Isuru Boyagane}
received his Bsc in Computer Science Engineering (Hons) from University of Moratuwa, Sri Lanka. He is currently a software engineer at WSO2. His research interests include Natural Language Processing, Information Retrieval and Distributed Systems.
\end{IEEEbiography}

\begin{IEEEbiography}
[{\includegraphics[width=1in,height=1.25in,clip,keepaspectratio]{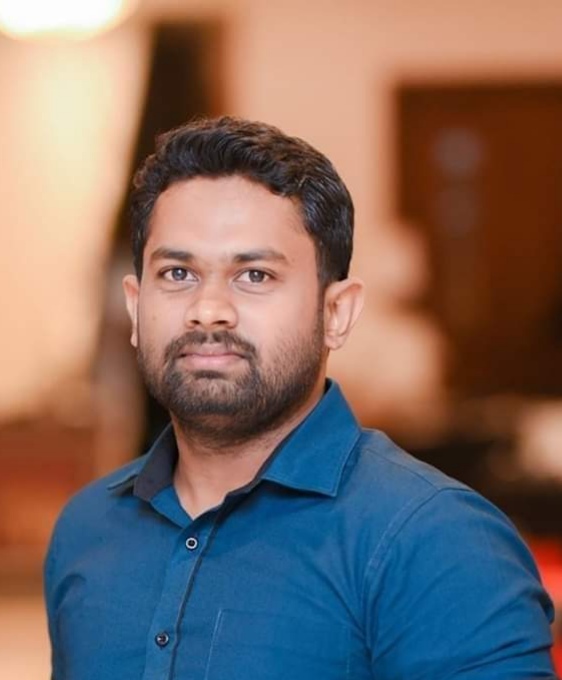}}]{Oshadha Katulanda}
received his Bsc in Engineering (Hons) in Computer Science from University of Moratuwa, Sri Lanka. He is currently a software engineer at 99x, Sri Lanka. His research interests include Natural Language Processing, Information Retrieval, and Machine Learning.
\end{IEEEbiography}


\begin{IEEEbiography}
[{\includegraphics[width=1in,height=1.25in,clip,keepaspectratio]{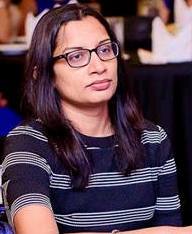}}]{Surangika Ranathunga}
received her Bsc in Engineering (Hons) and MSc in Computer Science from University of Moratuwa, Sri Lanka. She received her PhD from University of Otago, New Zealand. She is currently a senior lecturer at the Department of Computer Science and Engineering, University of Moratuwa, Sri Lanka. Her research interests include Natural Language Processing, Information Retrieval, and Machine Learning.
\end{IEEEbiography}

\begin{IEEEbiography}
[{\includegraphics[width=1in,height=1.25in,clip,keepaspectratio]{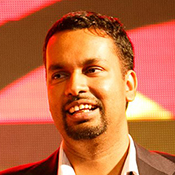}}]{Srinath Perera}
is the Chief Architect at WSO2,  a co-founder of the Apache Axis2 project,  and a member of the Apache Software Foundation. He has co-architected a dozen real-world distributed systems, including Apache Axis2, Apache Airvatha, and Siddhi Complex Event Processing Engine while authoring two books and 40+ technical articles.
\end{IEEEbiography}

\end{document}